\documentclass[preprint]{article}
\usepackage[left=2cm,right=2cm,top=2cm,bottom=2cm]{geometry}

\usepackage{authblk}
\usepackage{graphicx}
\usepackage{amsmath}
\usepackage[utf8]{inputenc}
\usepackage{hyperref}
\usepackage{url}
\usepackage{amssymb}
\usepackage{amsfonts}
\usepackage{indentfirst}
\usepackage[normalem]{ulem}
\usepackage[toc,title]{appendix}
\usepackage[font=small]{caption}
\usepackage{cite}
\usepackage{color}

\newcommand{\tr}{\textup{Tr}}
\newcommand{\btr}{\textup{bTr}}
\newcommand{\<}{\left<}
\renewcommand{\>}{\right>}
\newcommand{\idm}{\mathbf{1}}
\newcommand{\zb}{\bar{z}}
\newcommand{\wb}{\bar{w}}

\newcommand{\cG}{{\cal G}}
\newcommand{\cQ}{{\cal Q}}
\newcommand{\cX}{{\cal X}}

\newcommand{\cR}{{\cal R}}
\newcommand{\cB}{{\cal B}}
\newcommand{\ket}[1]{\left| #1 \>}
\newcommand{\bra}[1]{\< #1 \right|}
\newcommand{\braket}[2]{\< #1 | #2 \>}

\newcommand{\im}{\textup{Im}}
\newcommand{\be}{\begin{eqnarray}}
\newcommand{\ee}{\end{eqnarray}}

\begin{document}

\title{Complete diagrammatics  of the single ring theorem}
\author{Maciej A. Nowak\thanks{maciej.a.nowak@uj.edu.pl} }
\author{Wojciech Tarnowski\thanks{wojciech.tarnowski@student.uj.edu.pl}}
\affil{M. Smoluchowski Institute of Physics and
Mark Kac Complex Systems Research Center, Jagiellonian University,
S. Łojasiewicza 11,
PL 30-348 Kraków, Poland.}



\date{\today}
\maketitle
\begin{abstract}

Using diagrammatic techniques, we provide explicit functional relations between the cumulant generating functions for the biunitarily invariant ensembles in the limit of large size of matrices. The formalism allows to map two distinct areas of free random variables: Hermitian positive definite operators and non-normal R-diagonal operators. We also rederive the Haagerup-Larsen theorem and show how its recent extension to the eigenvector correlation function appears naturally within this approach.

\end{abstract}

\section{Introduction}
Over   half a century ago, Jean Ginibre~\cite{GINIBRE},  driven  solely by the  mathematical curiosity\footnote{In his own words: {\it Apart from the
intrinsic interest of the problem, one may hope that the methods and results will provide further insight in the cases of physical interest or suggest as yet lacking applications.}}, has considered the first non-normal random matrix model. He proposed   an ensemble, where the   elements of  a random  matrix were drawn from real  (complex, quaternion-valued) Gaussian distribution without imposing any restriction  on the symmetry of the ensemble.  Such simple random matrix model (named today as the Ginibre ensemble)  exhibited two main features  distinguishing it from earlier considered Hermitian  or unitary ensembles, which made a great impact in various fields of mathematics, physics,  statistics  and interdisciplinary applications.  First, the spectrum was complex, filling, in the limit when size of the matrix tends to infinity, a uniform disc on the complex plane. Second,  the ensemble is  non-normal, therefore possesses  two sets of left and right eigenvectors. It took several decades to understand the role of left and right eigenvectors.  In   late 90-ties, in a series of papers Fyodorov,  Savin, Sokolov,  Chalker and Mehlig~\cite{EIGENVECTORS}, have realized, that for non-normal matrices  another type of observables,  built out of right and left eigenvectors, plays  the crucial role in understanding non-Hermitian ensembles,  especially 
in the  study of stability of the ensemble under small perturbations. Nowadays, the non-normal random matrices turned out to be beneficial for the plethora of problems, involving chaotic scattering in quantum physics~\cite{CHAOTIC,RANDOMLASERS},  lagged cross correlations~\cite{BielyThurner,NowakTarnowskiLagged}, search algorithms~\cite{BACKSCATTERING}, non-Hermitian quantum mechanics~\cite{MOISEIEV,PTHATANO} and many others. 

From the perspective of this work, it is worthy of notice, that the Ginibre ensemble has another distinctive feature. 
The probability measure of the above-mentioned ensemble is invariant under the bi-unitary transformation, in contrast to single unitary invariance in the case of Dysonian threefold way~\cite{DYSONIAN}.  Since the unitary transformations related with the singular value decomposition fall into the symmetry group of the considered matrices, one expects that all spectral properties are given by the (squared) singular values. Symmetry of the ensemble assures also that the spectrum does not depend on the azimuthal angle $\phi$, but is a function of a radial variable $|z|$ only.

In the limiting case of the size of the matrix $N \rightarrow \infty$,  Ginibre ensemble was also the first case of called later {\bf R}-diagonal random matrices. The concept of {\bf R}-diagonal operator was  introduced formally  by Nica and Speicher~\cite{NicaSpeicherRDiagonal}, in the framework of Voiculescu's theory of free random variables~\cite{Voiculescu}.  We say that the operator  $X$ (or its matricial representation) is {\bf R}-diagonal,   if it  can  be decomposed as 
$X=PU$, where $P$ is Hermitian positive definite, $U$ represents Haar measure {\it and} $P$ and $U$ are mutually free. These are the natural extension of the isotropic complex random variables, the probability density function of which depends only of the modulus.  Since the  {\bf R}-diagonal operators  play a vital  role in several applications  of RMT - e.g. in MIMO telecommunication~\cite{MIMO} and quantum information problems~\cite{QUANTUM},  the study of {\bf R}-diagonal operators is not only an interesting subject from the viewpoint of formal mathematics. 

Perhaps the first  result for spectra of  biunitarily invariant matrices (but without explicit  relation to {\bf R}-diagonality) was formulated  in  a  paper by Feinberg and Zee~\cite{FeinbergZeeSingleRing}, who discovered the so-called "single ring theorem" - the spectrum of  the bi-unitary invariant  measure in the limit when the $N \rightarrow \infty$ is  always confined between two rings, $r_{in}$ and $r_{out}$. This theorem was elaborated later in more details by Feinberg, Scalettar and Zee in~\cite{FeinbergZeeSingleRing2}. The single ring theorem includes also the cases when $r_{in}=0$  (disc, alike Ginibre case) or  $r_{out}=\infty$.   Independently, in the more general framework of operator algebras Haagerup and Larsen~\cite{HaagerupLarsen}  have mathematically formulated the single ring theorem in terms of Voiculescu multiplicative S-transform~\cite{Stransform}    for the square of the polar operator $P$. Then the topic of {\bf R}-diagonal operators became the subject of intense research in mathematics.  Another proof relying on the characteristic determinant and integration over the unitary groups was given by 
Fyodorov and Khoruzenko~\cite{FyodorovKhoruzenko}. The direct and mathematically complete link to random matrices was established quite late by  Guionnet, Krishnapur and Zeitouini~\cite{GKZ}.   In 2015, Belinschi, Speicher and \'{S}niady~\cite{BSS}  have rigorously proven, how the single ring theorem emerges as a result of the reduction of "hermitization" ("quaternionization") approaches,  proposed in the context of physical problems involving non-Hermitian operators~\cite{USOLD,JaroszNowakNovel,OTHERS1,ChalkerWang,OTHERS2}.   All these works were concentrating  on spectral properties of the single ring theorem and did not address the issue of eigenvector correlations.  Very recently, the single ring theorem was augmented, using analytic methods,  with the part predicting also the generic form of a certain eigenvector correlation function~\cite{EgvCondNumb}. 

 Historically, Voiculescu formulated the Free Probability theory in the analytical language, but several years later, Speicher and Nica reformulated it as a combinatorial theory of non-crossing partitions, corresponding to planar diagrams in physicists language. 
Despite that the {\bf R}-diagonal concept was originally introduced in a combinatorial language,  all known mathematical proofs and generalization of the  Haagerup-Larsen theorem were performed using the analytic methods of Free Random Variables, sometimes quite involved.  Moreover, despite that the original Feinberg and Zee's approach to the single ring theorem relies on the resummation of planar diagrams, they do not focus in the their combinatorial aspects.

The intention of this work is to  fill this gap and provide  the simple  diagrammatic arguments leading directly to the Haagerup-Larsen theorem,  for both the spectra and  the left-right eigenvector correlator.  In this way we build {\it an explicit}  relation between the $R$-transform of the square of the Hermitian operator $P$ and the quaternionic  R-transform of {\bf R}-diagonal matrices. 

The paper is organized as follows. In  section~\ref{sec:Hermitian}, we recall the diagrammatics  of Hermitian ensembles  leading to the Voiculescu additive R-transform
for free convolution. We note here, that to avoid confusion coming from too many $R$'s used traditionally in the Free Random Variable calculus, we denote  the {\bf R}-diagonal  feature using the bold font.  Section~\ref{sec:NonHermitian} generalizes this diagrammatic construction to the case of non-Hermitian operators, following  quaternionization construction~\cite{USOLD,JaroszNowakNovel}.  Section~\ref{Main} shows the main result of the paper, i.e. the procedure of effective reduction of generic non-Hermitian diagrammatics  to the case of {\bf R}-diagonal operators.  The main result is the diagrammatic derivation of the full (spectra and eigenvectors) Haagerup-Larsen theorem.  We  also stress the analogy between the Hermitian  and {\bf R}-diagonal cases, by presenting the mapping between various transformations used in free probability.  We elucidate also  an  infinite resummation of the corresponding diagrams emerging from the change of the variables  and leading  to the change of the topology of the interlocked one-line irreducible diagrams. This observation is crucial for the proof. 
 Finally, in Section~\ref{sec:Applications} we consider  three explicit examples applying  our construction  for a simple rederivation of the quaternionic  R-transform for the Haar measure, the {\bf R}-diagonal analogue of free Poisson distribution and we study cumulants of the products of Ginibre matrices, which turn out to be the so-called Raney numbers. Section~\ref{sec:Summary} concludes the paper.

\section{Hermitian random matrices \label{sec:Hermitian}}
Before we focus on diagrams for non-Hermitian {\bf R}-diagonal matrices we 
 present briefly the diagrammatic approach to large Hermitian matrices and their integrable structure. We consider random matrices, the probability density function (pdf) of which is given by
\begin{equation}
P(H)dH=Z^{-1} \exp(-N\tr V(H)) dH. \label{eq:pdfHerm}
\end{equation}
Here $dH=\prod_{j=1}^{N}d\textup{Re} H_{jj}\prod\limits_{\substack{j,k=1 \\ j<k}}^{N}d\textup{Re}H_{jk}d\textup{Im}H_{jk}$ and $H_{jk}=\bar{H}_{kj}$. The potential $V(H)=\sum_{k=1}^{\infty} g_k H^k$ is chosen such that the pdf is normalizable. We also include a possibility that some of $g$'s are zero. The normalizing constant ensures that $\int P(H)dH=1$.  The pdf is invariant under the unitary transformations $H\to UHU^{\dagger}$, where $U\in U(N)$. 

\begin{figure}
\includegraphics[width=\textwidth]{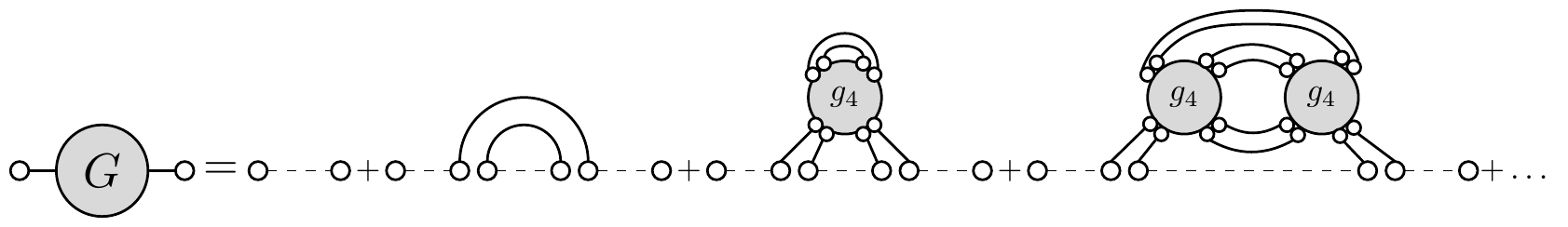}
\caption{Several exemplary planar diagrams which appear in the expansion of the complex resolvent for the potential $V(H)=g_2 H^2+g_4H^4$. Double dots denote a matrix $H$ and each dot carries an index (not shown explicitly) which refers to the entries of the matrix. A dashed horizontal line, with single dots at its ends represents $\frac{\idm}{z}$ terms in the expansion \eqref{eq:ResExp}. Gray circles denote the coefficients of the power expansion of the potential \eqref{eq:PotExp}, while budding white circles are the matrices in this expansion. For the simplicity, trivial summations of the identity matrix with the relevant matrices in the expansion are presented as already merged circles.
\label{Fig:DiagrExpansion}
}
\end{figure}

Since these  transformations can bring $H$ to the diagonal form, the only invariant quantities are built out of its eigenvalues. The simplest possible such object is the one-point correlation function, the spectral density
\begin{equation}
\rho(x):=\<\frac{1}{N}\sum_{j=1}^{N}\delta(x-\lambda_i)\>,
\end{equation}
where $\<\ldots\>$ denotes the average over the pdf \eqref{eq:pdfHerm}
\begin{equation}
\< f(H)\>=\int f(H) \exp(-N \tr V(H)) dH.
\end{equation}
In practice, the spectral density is inconvenient to handle directly, thus one considers its Stieltjes transform (Green's function)
\begin{equation}
G(z):=\<\frac{1}{N}\tr \frac{1}{z\idm - H}\>,
\end{equation}
which is more tractable. The spectral density can be recovered from $G$ by means of the Sochocki-Plemelj formula
\begin{equation}
\rho(x)=-\frac{1}{\pi} \lim_{\epsilon\to 0} \im G(x+i\epsilon).
\end{equation}
The Green's function is also the generating function for  moments of the spectral density, as can be seen from the power series expansion at $z=\infty$
\begin{equation}
G(z)=\sum_{k=0}^{\infty}\frac{1}{z^{k+1}}\<\frac{1}{N}\tr H^k\>=\sum_{k=0}^{\infty} \frac{1}{z^{k+1}}\int \rho(\lambda)\lambda^{k}d\lambda.
\end{equation}
For large matrices calculations of $G$  can be done perturbatively. Consider the averaged resolvent $\hat{G}=\<(z\idm-H)^{-1}\>$. The unitary invariance of the pdf asserts that any quantity calculated as an average of matrices $H$ has a trivial matrix structure, i.e. it is a multiple of the identity matrix (see \cite{Weingarten, Beenakker} for calculations of averages with respect to the unitary group and~\cite{Collins,CollinsSniady} for a modern approach).  We therefore treat them as scalars, which also enables us to freely interchange the resolvent with its traced version, the Green's function. The resolvent can be expanded into the geometric series
\begin{equation}
G \idm=\frac{\idm}{z}+\<\frac{\idm}{z}H \frac{\idm}{z}\>+\<\frac{\idm}{z}H \frac{\idm}{z}H \frac{\idm}{z}\>+\ldots  \label{eq:ResExp}
\end{equation}
and the average in each term can be evaluated independently. 
Sometimes it is convenient to rephrase the expansion into moments around $z=0$, then 
the useful generating function  reads
\be
\tilde{M}(z) := \frac{1}{z} G\left( \frac{1}{z} \right)\frac{1}{z} -\frac{1}{z} =\sum_{k=1}^{\infty}m_kz^{k-1}.
\label{Mbarherm}
\ee
For calculation of the averages in \eqref{eq:ResExp} we 
 use the diagrammatic representation, borrowing from standard field-theoretical tools -- Feynman diagrams. 
The main ingredient is the representation of the Kronecker delta function  by a single line. In the case of random matrix models the graphs 
have particularly simple form, since they basically control only the flow of the indices of the averaged strings of matrices. We distinguish the Gaussian part in the potential, while the remaining part of the potential we expand into a series
\begin{equation}
 \exp(-N\tr V(H))= \exp(-N g_2 \tr H^2)\sum_{l=0}^{\infty}\frac{1}{l!}\left(-N\sum_{k\neq 2}^{\infty} g_k H^k\right)^l. \label{eq:PotExp}
\end{equation}
Then we integrate it, term by term, with respect to the Gaussian measure, making use of the Wick's (Isserlis') theorem, which reduces the calculation of expectations of $H$'s to the all possible pairings. The price for the simplification of the integrals is the proliferation of the terms in the integrand. In order to cope with the multitude of expressions, we represent them graphically as diagrams, see Fig. \ref{Fig:DiagrExpansion}. A pairing of two matrices we call a propagator, which brings a factor
\begin{equation}
\<H_{ab}H_{cd}\>_G=\frac{1}{g_2 N}\delta_{ad}\delta_{bc}, \label{eq:propagator}
\end{equation}
which is represented by the double line. 
The terms on the r.h.s.  of \eqref{eq:PotExp} bring additional factors $Ng_k$ and $k$ matrices to be paired. The final expressions are represented by diagrams obtained from all possible pairing of vertices and matrices from the expansion \eqref{eq:ResExp} by the propagators \eqref{eq:propagator}. 

Remarkably, in the large $N$ limit only the planar diagrams give $\mathcal{O}(1)$ contribution, while the sum of all remaining diagrams vanishes in this limit, which simplifies the theory. This fact was observed by t'Hooft~\cite{tHooft} in the context of non-Abelian gauge theories.  We remark here that the subleading terms are succesively accessible in the framework of loop equations (see e.g. \cite{Eynard}) and the $\mathcal{O}(N^{-2g})$ corrections are encoded in diagrams that are planar when drawn on the surface of genus $g$.

\subsection{R-transform and 1LI diagrams}
Among the planar diagrams we distinguish a class of one-line irreducible (1LI) i.e. the ones that cannot be split into two by a single cut of a horizontal line. We denote by $\Sigma$ a sum of all 1LI diagrams and refer to this as self-energy. Further, among the 1LI diagrams we distinguish the connected subdiagrams which originate from the expectations $\<H^n\>$, where $H$'s are separated by dashed horizontal lines. The sum of such diagrams with the traced trivial matrix structure we call the $n$-th cumulant $\kappa_n:=\<\frac{1}{N}\tr H^n\>_c$ and endow the corresponding average with the  index $c$. We consider also a generating function of all cumulants $R(z):=\sum_{n=1}^{\infty}\kappa_n z^{n-1}$, called the R-transform.

\begin{figure}[h] \begin{center}
\includegraphics[width=0.7 \textwidth]{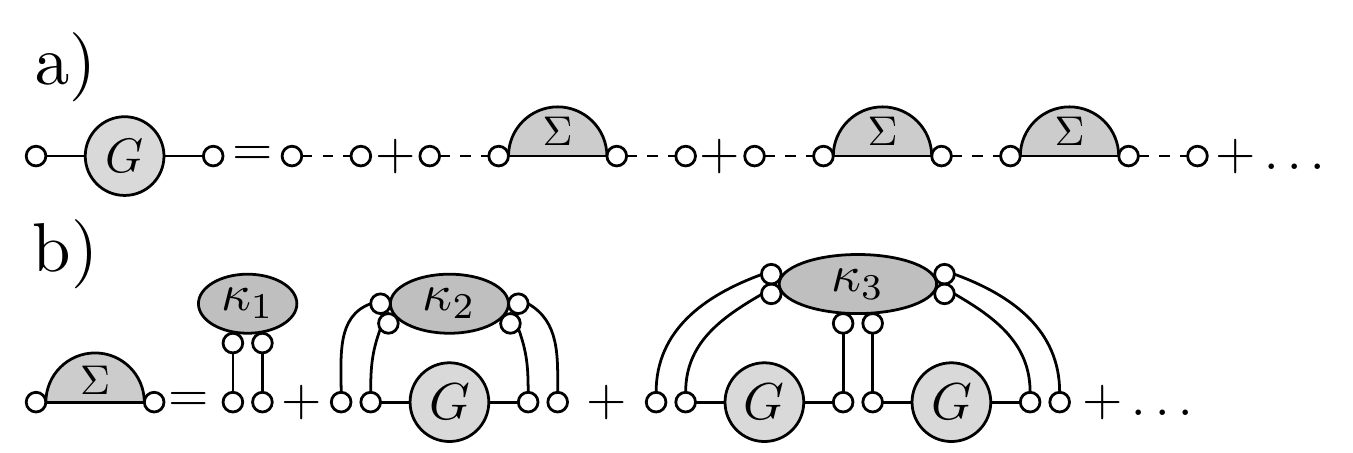}
\end{center}
\caption{a) Relation between the Green's function and the sum over all 1LI diagrams. b) Self energy represented by the cumulants and the Green's function. \label{Fig:SD}}
\end{figure}

The self-energy is the building block of the Green's function which can be expressed as a geometric series of $\Sigma$'s (see Fig. \ref{Fig:SD}a)
\begin{equation}
G(z)=\frac{1}{z}+\frac{1}{z}{\Sigma(z)}\frac{1}{z}+\frac{1}{z}{\Sigma(z)}\frac{1}{z}{\Sigma(z)}\frac{1}{z}+ \ldots,
\end{equation}
written in a closed form
\begin{equation}
G(z)=\left[z-\Sigma(z)\right]^{-1}.
\label{self}
\end{equation}
The relation between the Green's function and the self-energy is known under the name of Schwinger-Dyson equation. Moreover, the planarity  of diagrams allows also to express the self-energy through the Green's function by means of the $R$-transform (see Fig. \ref{Fig:SD}b)
\begin{equation}
\Sigma(z)=R(G(z))= \sum_{k=1}^{\infty} \kappa_k G^{k-1}(z)  \label{eq:SigmaR}
\end{equation}
Combining (\ref{self}) with (\ref{eq:SigmaR}) we arrive at the relation
\be
R[G(z)]+\frac{1}{G(z)}=z,
\label{Rfunction}
\ee
which, after introducing the auxiliary function $B(z)=R(z)+1/z$ (sometimes nicknamed Blue's function), leads to the  relation of the functional inverse type, i.e. 
\be
G[B(z)]=B[G(z)]=z.
\label{Blue}
\ee
Knowledge of the $R$-transform is sufficient to solve the matrix model in the large $N$ limit. We remark that the combinatorics standing behind the planar diagrams is equivalent to the axiomatic framework of the lattice of non-crossing partitions in the free probability theory~\cite{NicaSpeicher}.

\subsection{Addition and multiplication of Hermitian random matrices}

It is natural to ask a question about the spectral density of a sum of two large matrices. In general, this is a very hard task since the matrices can be correlated. In classical probability of real variables, one has to know their joint distribution in advance. The problem simplifies if they are independent and the logarithm of the Fourier transform of the probability density function is the quantity, which is additive when we add two independent random variables.

In the non-commutative world of large random matrices the notion of independence is replaced by freeness, a concept introduced by Voiculescu~\cite{Voiculescu}. We do not go into details concerning the free probability, we just mention an important fact that the matrices $A$ and $UBU^{\dagger}$, where $U$ is Haar unitary, are free in the limit when their size tends to infinity~\cite{VoiculescuAsymptotic,Xu}.
 In this framework, the $R$-transform, which has already appeared in the description of Feynman diagrams, is additive under the addition of free random variables. The additivity of the $R$-transform in the language of Feynman diagrams was shown in~\cite{ZEE}.
 
Although the product $AB$ of two Hermitian matrices is not Hermitian in general, assuming that at least one of these matrices, let us say $A$, is positive definite, the construction for multiplication is possible. Since $A$ is Hermitian positive definite, its square root $A^{1/2}$ is also Hermitian, therefore the matrix $AB$, isospectral to the Hermitian matrix $A^{1/2}BA^{1/2}$, has real eigenvalues. The problem of multiplying large unitarily invariant random matrices in the sense above is solved by the S-transform, which is related to the $R$-transform via
\begin{equation}
R(z)S(zR(z))=1, \qquad S(z)R(zS(z))=1. \label{eq:RelationRS}
\end{equation}
These relations are particularly simple, if we use "tilde-d"  functions  $\tilde{S}(z)= zS(z)$ and  $\tilde{R}(z)=zR(z)$. Then 
$\tilde{S}(z)$ is the functional inverse of $\tilde{R}(z)$
\be
\tilde{R}[\tilde{S}(z)]=\tilde{S}[\tilde{R}(z)]=z.
\ee
 This formulation requires $R_A(0)\neq 0\neq R_B(0)$, i.e. that the first moments of both distributions do not vanish, for the $S$-transform to exist. This assumption can be further weakened to the case when one of the distributions has zero mean~\cite{SpeicherRao}. There exists a different formulation of the multiplication rule in terms of the $R$-transforms solely~\cite{BJNProducts}
\begin{equation}
R_{AB}(z)=R_A(x)R_B(y),
\end{equation}
where $x$ and $y$ are related to $z$ via
\begin{equation}
x=zR_B(y),\quad y=zR_A(x).
\end{equation}

\section{Non-Hermitian random matrices \label{sec:NonHermitian}}
\subsection{Preliminaries}

In this section we focus on the spectral properties of non-Hermitian random matrices, the entries of which, $X_{jk}=x_{jk}+iy_{jk}$, are generated from the probability distribution
\begin{equation}
P(X,X^{\dagger})dXdX^{\dagger}\sim \exp\left[-N\tr V(X,X^{\dagger})\right]dXdX^{\dagger}, \label{eq:PotentialNonherm}
\end{equation}
where the measure reads
\begin{equation}
dXdX^{\dagger}=\prod_{j,k=1}^{N}dx_{jk}dy_{jk}.
\end{equation}
Since $X$ is not Hermitian, the potential depends also on $X^{\dagger}$ and non-normality of $X$ admits for much richer structure of the potential than in the Hermitian case. We demand that $V$ is chosen such that the probability density function is normalizable and real-valued. The pdf is again invariant under the unitary transformations. 

Non-Hermitian matrices have in general complex eigenvalues. In such a case the Sochocki-Plemelj formula ceases to work therefore in order to find a spectral density one has to resort to different methods. In the spirit of the electrostatic analogy one introduces a potential~\cite{ElectrostaticPotential,ElPot2,ElPot3}
\begin{equation}
\Phi(z,\zb,w,\wb):=\<\frac{1}{N}\log \det[ (z\idm-X)(\zb\idm-X^{\dagger})+|w|^2 \idm] \>
\label{FugKad}
\end{equation}
and its $z$-gradient
\begin{equation}
g(z,\zb,w,\wb):=\partial_z \Phi=\<\frac{1}{N}\tr \frac{\zb\idm-X^{\dagger}}{(z\idm-X)(\zb\idm-X^{\dagger})+|w|^2 \idm}\>. \label{eq:g}
\end{equation}
The spectral density $\rho(z,\zb):=\<\frac{1}{N}\sum_{i=1}^{N} \delta^{(2)}(z-\lambda_i)\>$ is then given by the Poisson law
\begin{equation}
\rho(z,\zb)=\frac{1}{\pi}\lim_{|w|\to 0} \partial_{\zb}g(z,\zb,w,\wb)=\lim_{|w|\to 0}\frac{1}{\pi}\partial_{z\zb}\Phi(z,\zb,w,\wb),
\end{equation}
which simply follows from the representation of the 2-dimensional Dirac delta (Poisson kernel)
\begin{equation}
\delta^{(2)}(z)=\lim_{\epsilon\to 0}\frac{1}{\pi}\frac{\epsilon^2}{(|z|^2+\epsilon^2)^2}.
\end{equation}

The $z$-gradient of the potential contains a quadratic expression in the denominator, which makes it inconvenient to calculate in the perturbative expansion. For the price of doubling the size of a considered matrix one can linearize $g$~\cite{USOLD,OTHERS1,ChalkerWang}, introducing the generalized Green's function, which is a $2\times 2$ matrix
\begin{equation}
\cG=\left(\begin{array}{cc}
\cG_{11} & \cG_{1\bar{1}} \\
\cG_{\bar{1}1} & \cG_{\bar{1}\bar{1}}
\end{array}\right):=\<\frac{1}{N}\btr \left(\begin{array}{cc}
z\idm-X & i\wb \idm \\
i w\idm & \zb\idm - X^{\dagger}
\end{array}\right)^{-1}\>,
\end{equation}
where we introduced a block trace operation, which, acting on $2N\times 2N$ matrices, yields a $2\times 2$ matrix
\begin{equation}
\btr\left(\begin{array}{cc}
A & B \\ 
C & D 
\end{array}\right)=\left(\begin{array}{cc}
\tr A & \tr B \\ 
\tr C & \tr D 
\end{array}\right).
\end{equation}
The components of the generalized Green's function read explicitly
\begin{equation}
\begin{array}{c}\cG_{11}=\<\frac{1}{N}\tr \frac{\zb\idm-X^{\dagger}}{(z\idm-X)(\zb\idm-X^{\dagger})+|w|^2\idm}\>=\bar{\cG}_{\bar{1}\bar{1}}, \\
\cG_{1\bar{1}}=\<\frac{1}{N}\tr \frac{-i\wb}{(z\idm-X)(\zb\idm-X^{\dagger})+|w|^2\idm}\>=-\bar{\cG}_{\bar{1}1}.
\end{array} \label{eq:GComponents}
\end{equation}

One easily recognizes that the $\cG_{11}$ entry of the generalized Green's function is exactly equal to \eqref{eq:g}. Moreover, the generalized Green's function can be written in the form which resembles the complex Green's function for Hermitian matrices
\begin{equation}
\cG(Q)=\<\frac{1}{N}\btr (Q \otimes \idm_N -\cX)^{-1}\>,
\end{equation}
where $Q$ is a $2\times 2$ matrix representation of a quaternion and $\cX$ is a duplicated matrix
\begin{equation}
Q:=\left(\begin{array}{cc}
z & i\wb \\
i w & \zb
\end{array} \right), \quad 
\cX:=\left(\begin{array}{cc}
X & 0 \\
0 & X^{\dagger}
\end{array}\right).
\end{equation}
The generalized Green's function as a $2\times 2$ matrix has a structure of a quaternion, therefore we refer to this also as the quaternionic Green's function~\cite{JaroszNowakNovel}.

Solving the random matrix problem for the spectral density, one always gets the entire generalized Green's function, obtaining the additional information stored in the off-diagonal elements of $\cG$, which is as important as the spectral density itself.

A non-Hermitian matrix, if it is diagonalizable, possesses sets of left ($\bra{L}$) and right ($\ket{R}$) eigenvectors, solving the eigenproblem
\begin{equation}
X\ket{R_i}=\lambda_i\ket{R_i}, \quad \bra{L_i}X=\bra{L_i}\lambda_i.
\end{equation}
The eigenvectors are biorthogonal, i.e. they are constrained by the condition $\braket{L_i}{R_j}=\delta_{ij}$, but they are not orthogonal among themselves $\braket{L_i}{L_j}\neq \delta_{ij}$. The bi-orthogonality condition leaves freedom of rescaling each eigenvector by a nonzero complex number $\ket{R_i}\to c_i\ket{R_i}$ with $\bra{L_i}\to \bra{L_i}c_{i}^{-1}$, and of multiplication by a unitary matrix $\ket{R_i}\to U\ket{R_i}$ with $\bra{L_i}\to\bra{L_i}U^{\dagger}$. 
 The vectors obtained by the latter transformation are no longer the eigenvectors of $X$, but of another matrix $UXU^{\dagger}$, which lies on the orbit of $X$ under the symmetry group of the joint pdf \eqref{eq:PotentialNonherm}.
The simplest non-trivial quantity invariant under these transformations is the matrix of overlaps $O_{\alpha\beta}:=\braket{L_{\alpha}}{L_{\beta}}\braket{R_{\beta}}{R_{\alpha}}$. Only for normal matrices $O_{ij}=\delta_{ij}$, therefore the overlaps can be thought of as a measure of non-normality of a matrix.

Chalker and Mehlig introduced in~\cite{ChalkerMehlig} a one-point correlation function associated with the diagonal part of the overlap matrix, being the special case of the Bell-Steinberger matrix~\cite{BellSteinberger}
\begin{equation}
O_{N}(z,\zb):=\<\frac{1}{N^2}\sum_{\alpha=1}^{N}O_{\alpha\alpha}\delta^{(2)}(z-\lambda_{\alpha})\>.
\end{equation}
In the large $N$ limit the eigenvector correlation function is given by the product of off-diagonal elements of the quaternionic Green's function\cite{JanikNowakCorr,EgvCondNumb}
\begin{equation}
O(z,\zb):=\lim_{N\to\infty} O_{N}(z,\zb)=-\frac{1}{\pi} \lim_{|w|\to 0} \lim_{N\to\infty} \cG_{1\bar{1}}\cG_{\bar{1}1}.
\end{equation}
The diagonal overlaps $O_{\alpha\alpha}$ are the squares of the eigenvalue condition numbers, $O_{\alpha\alpha}=\kappa^{2}(\lambda_{\alpha})$, known in the numerical analysis community to play the significant role in the stability of the spectrum against additive perturbations~\cite{Wilkinson,Trefethen}. The Cauchy-Schwartz inequality gives a bound $O_{ii}\geq 1$ and the inequality is saturated for normal matrices. The ratio $O_N(z,\zb)/\rho(z,\zb)$ gives a conditional expectation of the squared eigenvalue condition number~\cite{EgvCondNumb}
\begin{equation}
\frac{O_N(z,\zb)}{\rho(z,\zb)}=\mathbb{E}\left(\left. \frac{1}{N}\kappa^{2}(\lambda_{\alpha})\right| z=\lambda_{\alpha}\right).
\end{equation}

\subsection{Large N expansion of the quaternionic Green's function and Feynman diagrams}

The procedure for the calculation of the quaternionic Green's function is only slightly modified, compared to the Hermitian case. Again, the unitary invariance of the pdf asserts that the untraced resolvent has a trivial structure $\<(Q\otimes\idm-\cX)^{-1}\>=\cG\otimes \idm$. We write the geometric series
\begin{equation}
\cG(Q)\otimes\idm=\cQ^{-1}+\<\cQ^{-1}\cX\cQ^{-1}\>+\<\cQ^{-1}\cX\cQ^{-1}\cX\cQ^{-1}\>+\ldots,
\end{equation} 
where $\cQ=Q\otimes \idm$. Now, due to the block structure originating from the linearization, all objects in the above expansion, apart from the matrix indices, possess additional indices ($1,\bar{1}$) enumerating the blocks of $\cX$ and the elements of the quaternion. The block trace operation taken at the end of the calculations is in fact the partial trace over the matrix space.

To calculate the averages, we decompose the pdf into Gaussian and residual parts. The most general, allowed by the Hermiticity condition, Gaussian part of the potential \eqref{eq:PotentialNonherm} can be written in a convenient form with $\sigma>0$ and $\tau\in(-1,1)$
\begin{equation}
P_G(X,X^{\dagger})\sim \exp\left[-\frac{N}{\sigma^2} \frac{1}{1-\tau^2} \left(\tr XX^{\dagger}-\frac{\tau}{2} \tr (X^2+(X^{\dagger})^2)\right)\right].
\end{equation}
The propagator therefore reads
\begin{equation}
\<\cX_{ab}^{\alpha\beta}\cX_{cd}^{\mu\nu}\>_{G}=\frac{\sigma^2}{N}(1+(\tau-1)\delta_{\alpha\mu})\delta_{\alpha\beta}\delta_{\mu\nu}\delta_{ad}\delta_{bc},
\end{equation}
where the Greek indices take values from $\{1,\bar{1}\}$ and the Latin ones from $\{1,2,\ldots,N\}$. The residual part is expanded into a power series, bringing additional matrices (vertices in the diagrammatic representation), and all averages are then calculated with respect to the Gaussian measure, which by means of the Wick's theorem reduces to the summation over all possible pairings. The diagrammatic rules are exactly the same as in the Hermitian case, apart from the additional Greek indices, carried by each dot. In this paper we do not exploit the index structure explicitly, for a thorough calculation involving Gaussian matrices we refer to~\cite{NowakTarnowskiLagged}.

\subsection{Quaternionic R-transform}

The structure of the Feynman diagrams is exactly the same as for Hermitian matrices, but now the objects that we calculate are $2\times 2$ matrices. The Schwinger-Dyson equation relating the quaternionic Green's function with the self-energy composed of 1LI diagrams then reads
\begin{equation}
\cG(Q)=[Q-\Sigma(Q)]^{-1} .
\end{equation}
 Now, due to the fact that in general $X$ is not related with $X^{\dagger}$, there are many types of cumulants in the expansion of $\cG$, corresponding to the connected averages of different words, separated by a horizontal line (appropriate component of $Q^{-1}$), e.g. $c_{1\bar{1}1\bar{1}}=\<\frac{1}{N}\tr XX^{\dagger}XX^{\dagger}\>_c \neq \<\frac{1}{N}\tr XXX^{\dagger}X^{\dagger}\>_c=c_{11\bar{1}\bar{1}}$. Remarkably, all possible cumulants are
 stored in a single object, the quaternionic $R$-transform, which itself is a $2\times 2$ matrix representation of the quaternion, defined as follows
 \begin{equation}
 \cR(Q)\otimes \idm=\<\cX\>_c+\<\cX \cQ \cX\>_c+\<\cX \cQ\cX\cQ\cX\>_c+\ldots, \label{eq:Rexp}
 \end{equation}
 more explicitly
 \begin{equation}
 \cR(Q)_{\alpha\beta}=c^{(1)}_{\alpha} \delta_{\alpha\beta}+c_{\alpha\beta}^{(2)} Q_{\alpha\beta}+\sum_{\gamma\in \{1,\bar{1}\} }c_{\alpha\gamma\beta}^{(3)} Q_{\alpha\gamma}Q_{\gamma\beta}+\sum_{\gamma,\epsilon \in \{1,\bar{1}\} }c_{\alpha\gamma\epsilon\beta}^{(4)} Q_{\alpha\gamma}Q_{\gamma\epsilon}Q_{\epsilon\beta}+\ldots\,\, . \label{eq:Rtransform}
 \end{equation}
 
 This definition is quite compact and  deserves more  intuitive explanation. Suppose that we know  all cumulants and we want to construct the $\cR$-transform. We naturally associate $\bar{1}$ in the index of the cumulant with $\dagger$ in the corresponding expression in $X$'s. The first and the last index of the cumulant give us the appropriate component of $\cR$. Starting from the first index, we move towards the rightmost one and each time we make a step between two indices, we pick the component of $Q$ given by the indices we encounter. $Q$ therefore can be considered as a transfer matrix. The cumulant $c^{(4)}_{\alpha\beta\gamma\epsilon}=\<\frac{1}{N} \tr X^{\alpha}X^{\beta}X^{\gamma}X^{\epsilon}\>_c$ comes with the expression $Q_{\alpha\beta}Q_{\beta\gamma}Q_{\gamma\epsilon}$ in $\cR_{\alpha\beta}$. For example,  $c_{111\bar{1}}$ appears with $Q_{11}Q_{11}Q_{1\bar{1}}=z^2(i\wb)$ in $\cR_{1\bar{1}}$. The ability to store all mixed cumulant in a single object relies on the fact that $\cQ $ and $\cX$ do not commute. The mapping between $\cR$-transform and the cumulants is not bijective, there are different cumulants, which bring the same expression in the components of $Q$ to the quaternionic $R$-transform\footnote{For example a pair $c^{(6)}_{11\bar{1} 1\bar{1} 1} $ and $c^{(6)}_{1\bar{1} 11\bar{1} 1}$}. For the one-to-one mapping, one has to consider either different $Q$'s in the expansion \eqref{eq:Rexp} or a single $Q$, but with entries from a noncommutative algebra.

The relation between the self-energy and the quaternionic Green's function through the connected diagrams can be expressed via the quaternionic $R$ transform $\Sigma(Q)=\cR(\cG(Q))$. A direct relation between the generalized Green's function and the quaternionic $R$-transform can be written in terms of the auxiliary funcion, nicknamed Blue's function, which is the functional inverse of the quaternionic Green's function $\cB(\cG(Q))=Q=\cG(\cB(Q))$. The $\cR$-transform is then given by $\cR(Q)=\cB(Q)-Q^{-1}$. The inverse of $Q$ is understood as the matrix inverse.

The $R$-transform for non-Hermitian matrices was discovered in~\cite{USOLD,FeinbergZeeSingleRing} as a function generating all 1LI diagrams. The quaternionic structure was discovered much later~\cite{JaroszNowakNovel}.   

We remark here that the mixed moments are encoded in the same way in the quaternionic moment generating function
\begin{equation}
\tilde{{\cal M}}(Q)=Q^{-1}\cG(Q^{-1})Q^{-1}-Q^{-1}.
\end{equation}

We conclude this section by a comparison of two formalisms (Table~\ref{Tab:Comparison}),  using a calligraphic notation in the case of non-Hermitian analogues of Hermitian entries.  

\begin{table}[h]
\begin{center}
{\small
 \begin{tabular}{||c c c ||} 
 \hline
  & Hermitian   & Non-Hermitian   \\ 
  \hline\hline
 Spectrum  & real  & complex  \\ 
 \hline
 Green's   & complex-valued & quaternion-valued  \\
 
 function & $G(z)=\frac{1}{N} \left< {\rm Tr} (z-H)^{-1}  \right>$ \,\,&   ${\cal G}(Q)=\frac{1}{N} \left< {\rm bTr} (Q-{\cal X})^{-1}  \right>$\\
 \hline
Moments  & $\tilde{M}(z)=\frac{1}{z}G\left(\frac{1}{z}\right) \frac{1}{z} -\frac{1}{z}$\,\,& $ \tilde{{\cal M}}(Q)=Q^{-1}\cG(Q^{-1})Q^{-1}-Q^{-1}$ \\
 \hline
Cumulants  & $R(z)=\sum_{n} \kappa_n z^{n-1}$\,\,   & $[{\cal{R}}(Q)]_{\alpha \beta}=\sum_{k,\{ i_1...i_{k-2} \} } c^{(k)}_{\alpha i_1...i_{k-2} \beta} Q_{\alpha i_1}...Q_{i_{k-2} \beta}$\\ 
 \hline
S-D eqs.  &    $ R[G(z)]+\frac{1}{G(z)}=z$         & $ {\cal{R}}[{\cal{G}}(Q)] +[{\cal{G}}(Q)]^{-1} =Q$ \\
  \hline
 \hline 
\end{tabular}
}
\end{center}
\caption{Comparison between corresponding quantities in Hermitian versus non-Hermitian ensembles.\label{Tab:Comparison}} 
\end{table}

In the next chapter we demonstrate, how, in the case of {\bf R}-diagonal operators,  the general formalism for non-Hermitian ensembles reduces to quasi-Hermitian  formalism    for bi-unitary invariant ensembles. 

\section{Bi-unitarily invariant random matrices \label{Main}}
The presented construction for non-Hermitian random matrices, despite its neatness, suffers from a limited practical usage due to the procedure of functional inversion on route from the $R$-transform to the quaternionic Green's function. Even in the calculation for Hermitian matrices we are limited by the degree of the polynomial equation that we have to solve.

There are classes of random matrices, the quaternionic $R$-transform of which can be calculated from simpler objects, because the spectrum of such matrices is one dimensional and matrices are normal. An embedding of complex transforms of Hermitian matrices to the general setting of quaternionic transforms in the non-Hermitian world was rendered by ~\cite{JaroszNowakNovel}. Later, an analogous embedding has been conducted for unitary matrices~\cite{JaroszUnitary}. 

In this section we discuss another class of random matrices, the spectrum of which, despite being complex, is effectively one dimensional, because the spectral problem has an azimuthal symmetry.  We consider matrices generated according to the probability distribution function $P(X,X^{\dagger})\sim \exp(-N\tr V(XX^{\dagger}))$. The  symmetry in this case is enhanced from $U(N)$ to $U(N)\times U(N)$.  The spectrum of $X$ is rotationally symmetric on the complex plane $\rho(z,\zb)=\rho_{r}(|z|)$ and the entire information is encoded in the radial cumulative distribution function $F(s)=\int_{|z|\leq s}\rho(z,\zb)d^2z=2\pi \int_{0}^{s}s'\rho_r(s')ds'$. Such matrices are the natural extensions of the so called isotropic complex random variables, the distribution of which depends only on their modulus.
Moreover, the symmetry  transformations can bring the matrix to the diagonal form with the singular values on the diagonal. It is natural, therefore, to expect all spectral properties of $X$ to be determined by its singular values. 

In the free probability community such objects are called {\bf R}-diagonal, the meaning of this notion shall become clear later. The {\bf R}-diagonal operators have a polar decomposition $X=PU$, where $P$ is Hermitian positive definite, $U$ is Haar unitary and $P$ and $U$ are mutually free. In the limit $N\to\infty$ biunitarily invariant random matrices become the $R$-diagonal operators, which was shown by Hiai and Petz~\cite[Theorem 4.4.5]{HiaiPetz}.

The first relation between the Green's function of $XX^{\dagger}$, encoding the distribution of the squares of the singular values, and the radial cumulative distribution was found by Feinberg and Zee~\cite{FeinbergZeeSingleRing}. They also found a very intriguing property that the support of the spectrum of such matrix is either a disc or an annulus, which bears the name of the single ring theorem. Later, Haagerup and Larsen~\cite{HaagerupLarsen} within the framework of free probability derived a simple relation between the S-transform of $XX^{\dagger}$ and the radial cumulative distribution function
\begin{equation}
S_{XX^{\dagger}}(F(s)-1)=\frac{1}{s^2} \label{eq:HaagLars}
\end{equation}
Recently~\cite{EgvCondNumb}, this theorem has been extended to describe also the eigenvector correlation function
\begin{equation}
O(s)=\frac{F(s)(1-F(s))}{\pi s^2}. \label{eq:HaagLarsVect}
\end{equation}
The relation between eigenvalues and squared singular values of biunitarily invariant random matrices has been pushed much further by Kieburg and K\"osters who found an explicit integral transform between their joint pdfs~\cite{KieburgKosters}.

We remark here that the expression on the right-hand-side of the equality above has already appeared in the pioneering paper by Feinberg and Zee~\cite{FeinbergZeeSingleRing}, but without any connection to eigenvectors.

\subsection{1PI diagrams}

Due to the particular form of the potential, namely that $X$ and $X^{\dagger}$ appear alternately, and the planarity of diagrams in the large $N$ limit, the structure of diagrams simplifies considerably. 

The only non-vanishing in the large $N$ limit connected diagrams correspond to the expressions where $X$ and $X^{\dagger}$ are alternating, i.e. $\alpha_n:=\<\frac{1}{N}\tr (XX^{\dagger})^n\>_c=c^{(2n)}_{1\bar{1}1\bar{1}\ldots 1\bar{1}}=c^{(2n)}_{\bar{1}1\bar{1}1\ldots \bar{1}1}$. Let us consider a generating function for all such cumulants $A(x):=\sum_{k=1}^{\infty}\alpha_k z^{k-1}$. In the free probability context it is also known as the determining sequence~\cite{NicaSpeicherRDiagonal}.   Due to the very simple structure of non-vanishing cumulants, only one (commuting) variable is sufficient to encode all cumulants.  

According to the prescription \eqref{eq:Rtransform}, the quaternionic $R$-transform for biunitarily invariant random matrices  reads
\begin{equation}
\cR=A(-|w|^2) \left(\begin{array}{cc}
0 & i\wb \\
i w & 0
\end{array}\right). \label{eq:QuatRTransform}
\end{equation}
The $R$-transform depends only on the off-diagonal elements of the quaternion, which is a consequence of the fact that $X^{\dagger}$ has to be sandwiched between $X$'s and vice versa. In the 'hermitization' approach to non-Hermitian matrices, the corresponding $R$-transform is diagonal.

\subsection{Relation between R-transform of $XX^{\dagger}$ and the quaternionic R-transform of $X$}

Before we find a relation between the $R$-transforms, let us briefly explain why the alternating cumulants in the expansion of the quaternionic Green's function of $X$ are not the same as the corresponding ones of $XX^{\dagger}$. In both cases we consider moments like $\<\frac{1}{N}\tr (X^{\dagger}X)^k\>$ in the moment expansion of the Green's function of $XX^{\dagger}$. The average is taken over the probability measure proportional to $\exp(-N\tr V(X^{\dagger}X))dXdX^{\dagger}$. There are two ways of calculating such an object.

 First, making use of the symmetry of the potential, one changes the integration measure from $dXdX^{\dagger}$ into $d(XX^{\dagger})$ and uses the tools for Hermitian matrices. In this approach, however, the resulting Jacobian modifies the form of the potential, which changes the structure of the vertices in the expansion of the Green's function. 

In the diagrammatic approach for non-Hermitian problems, we circumvent the calculation of the Jacobian, calculating the averages with respect to the original measure. We remind that the connected diagrams contributing to the $k$-th cumulant originate from the $k$ matrices separated by horizontal lines. In the expansion of the Green's function of $XX^{\dagger}$ the matrices $X$ and $X^{\dagger}$ are merged and treated as a new single object, which gives rise to the new connected diagrams. Some propagators become internal lines, changing the topology of the diagrams and producing effectively new types of vertices. The notion of 1LI diagrams is always tied to the Green's function, the expansion of which is taken under consideration.

\begin{figure}
\begin{center}
\includegraphics[width=0.5\textwidth]{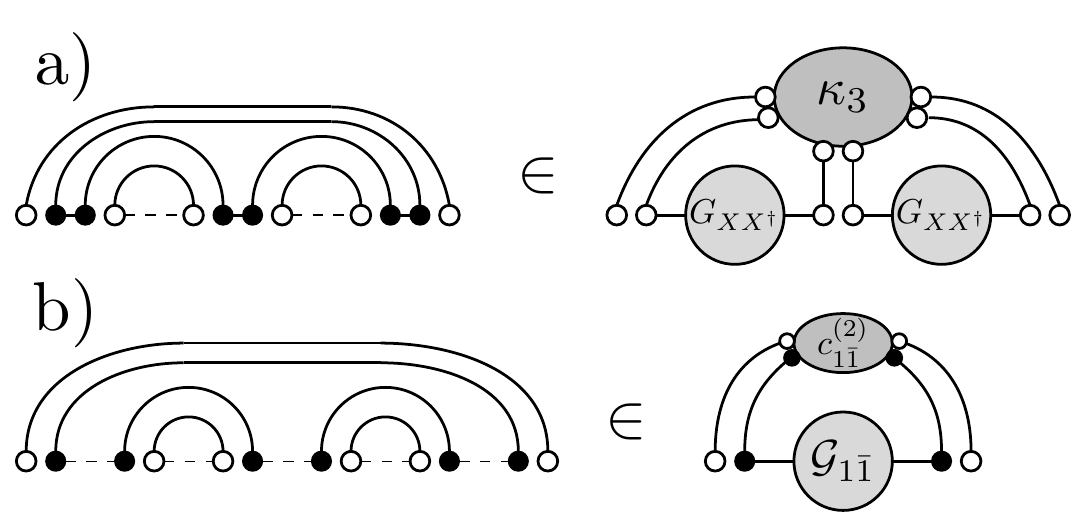}
\end{center}
\caption{The average $\< (XX^{\dagger})^{3}\>$ with respect to the Gaussian measure $\exp(-N\tr XX^{\dagger})$ produces the same vertices which are classified to different type of 1LI diagrams, depending on at which expansion it appears. Diagram a) appears in the expansion of the complex resolvent of $XX^{\dagger}$, while b) in the expansion of the quaternionic Green's function of $X$. To distinguish between $X$ and $X^{\dagger}$ we denote the first by double dots with white on the left, while for $X^{\dagger}$ the other way around. Two neighboring black dots joined by a solid line denote matrix multiplication.  The loop formed from the black dots and propagators in a) becomes an internal line of the cumulant $\kappa_3$ of $XX^{\dagger}$. The appearance of additional horizontal dashed lines (appropriate components of $Q^{-1}$) in b) classifies this diagram to a different class of 1LI diagrams.
 \label{Fig:Jacobian}
}
\end{figure}

In Fig. \ref{Fig:Jacobian} we present how the third cumulant of $XX^{\dagger}$ emerges from propagators, which are the only allowed lines for the quadratic potential. Such a mechanism is at the heart of the powerful linearization technique, which allows for calculation of spectra of products~\cite{BurdaJanikWaclaw,BJNProducts,MatrixDiffusion,GaussianProducts}, polynomials~\cite{NoncommutativePolynomials,NoncommutativePolynomials2} and even of the rational expressions of random matrices~\cite{NoncommutativeRational}.


Let us consider the expansion of the complex resolvent of $XX^{\dagger}$
\begin{equation}
G_{XX^{\dagger}}(z)\idm=\frac{\idm}{z}+\<\frac{\idm}{z}XX^{\dagger}\frac{\idm}{z}\>+\<\frac{\idm}{z}XX^{\dagger}\frac{\idm}{z}XX^{\dagger}\frac{\idm}{z}\>+\ldots\,\, .
\end{equation}

\begin{figure}
\includegraphics[width=\textwidth]{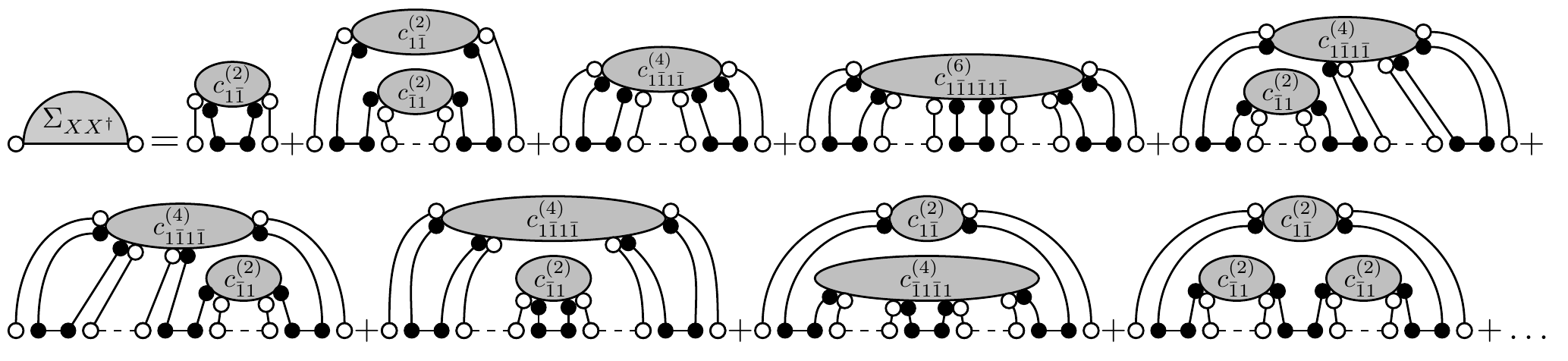}
\caption{Several lowest order 1PI diagrams contributing  to $\Sigma_{XX^{\dagger}}$.  \label{Fig:1PIresolvent}}
\end{figure}

\begin{figure}
\includegraphics[width=\textwidth]{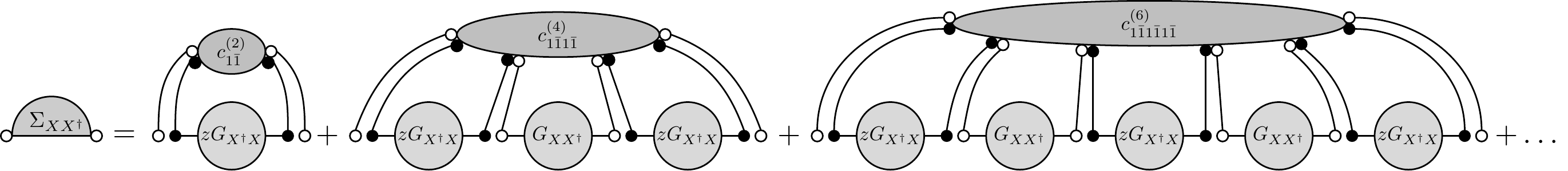}
\caption{General structure of 1PI diagrams in the resolvent expansion of $XX^{\dagger}$. \label{Fig:1PIGeneralStructure} }
\end{figure}

Let us focus on 1PI diagrams, the simplest of which are presented in Fig. \ref{Fig:1PIresolvent}. All vertices are already summed, such that we depicted cumulants. Let us notice that in order to have a 1LI diagram, the leftmost $X$  has to be connected with the rightmost $X^{\dagger}$ through some cumulant. The matrices in between legs of the outermost cumulant can be connected in any manner. Further, we use the fact that the term $\frac{\idm}{z}$ commutes with $X$ and $X^{\dagger}$, therefore we can rearrange diagrams, so that the resulting diagrams in between legs are the ones of the Green's function of either $XX^{\dagger}$ or $X^{\dagger}X$. The missing $\idm/z$ terms are encapsulated by considering $zG(z)$. The general structure of the diagrams, presented graphically in Fig. \ref{Fig:1PIGeneralStructure}, is now clear. The equation generated by these diagrams reads
\begin{eqnarray}
\Sigma_{XX^{\dagger}}(z)=c_{1\bar{1}}^{(2)}zG_{X^{\dagger}X}(z)+c^{(4)}_{1\bar{1}1\bar{1}}zG_{X^{\dagger}X}(z)G_{XX^{\dagger}}(z)zG_{X^{\dagger}X}(z)+\ldots = \\
zG_{X^{\dagger}X}(z)\sum_{k=1}^{\infty}\alpha_k(zG_{X^{\dagger}X}(z)G_{XX^{\dagger}}(z))^{k-1}=zG_{X^{\dagger}X}(z)A\left(zG_{X^{\dagger}X}(z)G_{XX^{\dagger}}(z)\right).
\end{eqnarray}

Using the general relation between the self-energy and the Green's function \eqref{eq:SigmaR}, we arrive at
\begin{equation}
R_{XX^{\dagger}}(G_{XX^{\dagger}}(z))=zG_{X^{\dagger}X}(z)A\left(zG_{X^{\dagger}X}(z)G_{XX^{\dagger}}(z)\right).
\end{equation}
Knowing that for square matrices $G_{XX^{\dagger}}=G_{X^{\dagger}X}$, which follows from cyclicity of the trace, we can associate these object and skip the subscript for simplicity. Let us also make a substitution $z\to B(z)$ and make use of the fact that $B$ is a functional inverse of $G$, to finally obtain
\begin{equation}
R(z)=zB(z)A\left(z^2B(z)\right). \label{eq:ARrel}
\end{equation}
This relation is convenient if knowing the quaternionic $R$-transform of $X$ one wants to calculate the complex $R$-transform of $XX^{\dagger}$. To invert this relation, let us introduce an auxiliary variable $y:=z^2 B(z)$. Note that $zR(z)+1=zB(z)=yA(y)+1$.


 
 Introducing yet another variable $t$ given by $z=:tS(t)$, where $S(z)$ is the $S$ transform and using the relation between $R$ and $S$ transforms \eqref{eq:RelationRS}, we obtain 
\begin{equation}
t=yA(y)\quad \mbox{and} \quad y=t(t+1)S(t). \label{eq:aux1}
\end{equation}
Let us introduce another auxiliary transform $K(z)$ related to $A$ in a similar way as $R$ is related with $S$, namely
\begin{equation}
A(z)K(zA(z))=1, \quad K(z)A(zK(z))=1. \label{eq:AKrelation}
\end{equation}
This definition says that $zK(z)$ is a functional inverse of $zA(z)$. Such an inversion is always possible for non-zero {\bf R}-diagonal operators, since $A(0)=c_{1\bar{1}}^{(2)}=\< \frac{1}{N}\tr XX^{\dagger}\>$. 
The last auxiliary variable $u$, which we define via $y=:uK(u)$,  transfers \eqref{eq:aux1} into
\begin{equation}
t=u \quad \mbox{and} \quad uK(u)=t(t+1)S(t).
\end{equation}
This gives us the simple relation between $K$ and $S$
\begin{equation}
S(t)=\frac{1}{1+t}K(t).
\end{equation}
The $S$ transform of $XX^{\dagger}$ is therefore related to the determining sequence of $X$ via
\begin{equation}
S(zA(z))=\frac{1}{A(z)(1+zA(z))}. \label{eq:SArelation}
\end{equation}
This relation is crucial in the derivation of the Haagerup-Larsen theorem, as we demonstrate in the next chapter. One can apply the relation between $R$ and $S$ transforms \eqref{eq:RelationRS} to obtain an equation which allows to calculate $A$ from $R$
\begin{equation}
R\left(\frac{z}{1+zA(z)}\right)=(1+zA(z))A(z). \label{eq:RArel}
\end{equation}
We remark that the relations \eqref{eq:ARrel} and \eqref{eq:RArel} were known earlier in terms of coefficients of their expansions around $z=0$~\cite[Proposition 15.6]{NicaSpeicher}. The  functional relation, which is more operational, and all intermediate steps are, to the best of our knowledge, the new results.

\begin{figure}
\includegraphics[width=\textwidth]{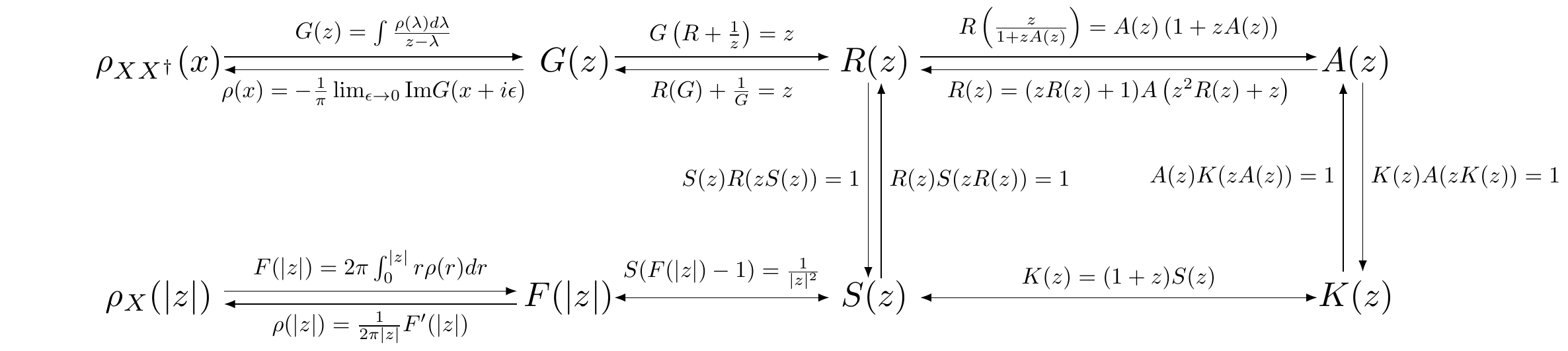}
\caption{Diagram of intermediate transforms, linking the spectral density of squared singular values and the radial profile of the eigenvalue density. The diagram also present the relation and all intermediate steps between the $R$-transform of squared singular values and the determining sequence of an {\bf R}-diagonal operator. The lower left branch is discussed in Sec. \ref{sec:HL}. 
 \label{Fig:Diagram} }
\end{figure}

To summarize this part, we have introduced a set of auxiliary transforms in order to relate the $R$-transform of $XX^{\dagger}$ with the determining sequence $A(z)$, being the cornerstone of the quaternionic $R$-transform of {\bf R}-diagonal matrices. In Fig. \ref{Fig:Diagram} we provide a guide through all transformations.

\subsection{The Haagerup-Larsen theorem \label{sec:HL}}

Using the above formalism we rederive in  a compact way the equations \eqref{eq:HaagLars} and \eqref{eq:HaagLarsVect}. 

Making use of the form of the quaternionic $R$-transform \eqref{eq:QuatRTransform} and the relation $\cR(\cG)+\cG^{-1}=\cB(\cG)=Q$, we obtain the matrix equation
\begin{equation}
A(\cG_{1\bar{1}}\cG_{\bar{1}1})\left(\begin{array}{cc}
0 & \cG_{1\bar{1}} \\
\cG_{\bar{1}1} & 0 
\end{array}\right)+\frac{1}{\cG_{11}\cG_{\bar{1}\bar{1}}-\cG_{1\bar{1}}\cG_{\bar{1}1}}\left(\begin{array}{cc}
\cG_{\bar{1}\bar{1}} & -\cG_{1\bar{1}} \\
-\cG_{\bar{1}1} & \cG_{11}
\end{array}\right)=\left(\begin{array}{cc}
z & i\wb \\
iw & \zb 
\end{array}\right).
\label{mainequationHL}
\end{equation}
We are now interested in calculating the spectral density and the eigenvector correlator, so we set $|w|\to 0$. Let us first consider the upper-left component of matricial eq.~(\ref{mainequationHL}):
\begin{equation}
\frac{\cG_{\bar{1}\bar{1}}}{\cG_{11}\cG_{\bar{1}\bar{1}}-\cG_{1\bar{1}}\cG_{\bar{1}1}}=z. \label{eq:HL11}
\end{equation}
The $\bar{1}\bar{1}$ component gives the complex conjugate of the above. Combining them together we easily deduce that $z\cG_{11}=\zb\cG_{\bar{1}\bar{1}}$. Denoting $F:=z\cG_{11}$~\cite{FeinbergZeeSingleRing}, we immediately obtain from \eqref{eq:HL11} that 
\begin{equation}
O(z,\zb)=-\frac{1}{\pi}\cG_{1\bar{1}}\cG_{\bar{1}1}=\frac{F-F^2}{\pi |z|^2}. \label{eq:CorrHL}
\end{equation}

Considering now the $1\bar{1}$ component we obtain two possibilities. First $\cG_{1\bar{1}}=0$ and in consequence $\cG_{11}=\frac{1}{z}$, which is a trivial solution, valid outside the spectrum.  Second, assuming $\cG_{1\bar{1}}\neq 0$ we  use \eqref{eq:CorrHL} to arrive at
\begin{equation}
A(\cG_{1\bar{1}}\cG_{\bar{1}1})\cG_{1\bar{1}}\cG_{\bar{1}1}=F-1.
\end{equation}
Now, evaluating the $S$ transform at both sides of the equation above, exploiting the relation between $S$ and $A$ \eqref{eq:SArelation}, and using \eqref{eq:CorrHL} once again, we finally get
\begin{equation}
S(F-1)=\frac{1}{A(\cG_{1\bar{1}}\cG_{\bar{1}1})(1+\cG_{1\bar{1}}\cG_{\bar{1}1}A(\cG_{1\bar{1}}\cG_{\bar{1}1}))}=\frac{1}{|z|^2}, \label{eq:HaagLars2}
\end{equation} 
which is the statement of the original formulation of the Haagerup-Larsen theorem. One deduces that $F$ depends on $z$ and $\zb$ only through their modulus, therefore the spectral density can be calculated by~\cite{FeinbergZeeSingleRing}
\begin{equation}
\rho(z,\zb)=\frac{1}{\pi} \partial_{\zb}\cG_{11}=\frac{1}{2\pi |z|}F'(|z|).
\end{equation}
Moreover, outside the support of the spectral density, the trivial solution gives $F=1$, therefore $F$ is indeed the radial cumulative distribution function. The inner and outer radii of the spectrum can be calculated by imposing $F=1$ (outer) or that $F$ is equal to the fraction of the zero modes (inner) and solving the resulting equation for $|z|$. In general, \eqref{eq:HaagLars2} can yield several solutions for $F$. The uniqueness of the radial cumulative distribution function has been shown within the framework of the analytic subordination function theory~\cite{BSS}.

Multiplying \eqref{eq:HaagLars2} by $(F-1)$, evaluating the $R$ transform on both sides of the equation and making use of \eqref{eq:RelationRS}, we obtain
\begin{equation}
R_{XX^{\dagger}}\left(\frac{F-1}{r^2}\right)=r^2.
\end{equation}
Substituting $F=1$ and taking into account that $R(0)=\kappa_1=m_1$, we relate the external radius with the first moment of $XX^{\dagger}$. 

Using the relation $S_{X}(z)S_{X^{-1}}(-1-z)=1$~\cite{HaagerupSchultz} and performing analogous computations, we arrive at
\begin{equation}
R_{(XX^{\dagger})^{-1}}(-Fr^2)=\frac{1}{r^2}.
\end{equation}
Substitution $F=0$ relates the first inverse moment of $XX^{\dagger}$ with the internal radius. To summarize, the internal and external radii are given by
\begin{equation}
r_{ext}^2=\int x \rho_{XX^{\dagger}}(x)dx, \qquad r_{in}^{-2}=\int x^{-1}\rho_{XX^{\dagger}}(x)dx,
\end{equation}
which was observed independently in~\cite{FeinbergZeeSingleRing2,HaagerupLarsen}. 

\subsection{Addition and multiplication of {\bf R}-diagonal matrices}
The problem of addition of unitarily invariant non-Hermitian random matrices was posed and solved long time ago~\cite{USOLD,FeinbergZeeSingleRing}. It turns out that the quaternionic $R$-transform is additive under the addition of non-Hermitian matrices, generalizing the result from free probability. Due to the particular form of the quaternionic $R$-transform for biunitarily invariant large random matrices {\bf R}-diagonal operators, their addition boils down to the addition of the corresponding determining sequences.

The multiplication is also straightforward for the reason that the entire one-point spectral information is encoded in the $S$-transform of $XX^{\dagger}$, by virtue of the Haagerup-Larsen theorem. Yet, the matrix $ABB^{\dagger}A^{\dagger}$ has the same eigenvalues as $BB^{\dagger}A^{\dagger}A$, and if $A$ and $B$ are free, then $S_{ABB^{\dagger}A^{\dagger}}(z)=S_{A^{\dagger}A}(z)S_{BB^{\dagger}}(z)$, which shows that the $S$ transform of $XX^{\dagger}$ is multiplicative.

We remark here that the rules for multiplication and addition of {\bf R}-diagonal operators were known in free probability, but were given in terms of a boxed convolution of determining sequences with an auxiliary M\"obius sequence~\cite{NicaSpeicherRDiagonal}. Our approach, which uses functions instead of the coefficients of their expansion is more operational. Together with the Haagerup-Larsen theorem this allows to calculate spectral densities and the eigenvector correlator easily. The presented  functional approach is complementary to the special case of the rectangular free probability~\cite{BenaychGeorges}. Within the latter formalism, when adding {\bf R}-diagonal operators, one deals with a symmetrized distribution of $X^{\dagger}X$, while our approach provides a very simple prescription for calculation of an additive quantity. Moreover, the explicit form of the quaternionic $R$-transform enables us also for addition with non-Hermitian random matrices which are not biunitarily invariant.  

\subsection{Abelization}

One way of generating complex isotropic variables is given in terms of the radial profile of the corresponding pdf. Its natural extension for noncommutative random variables is given by the Haagerup-Larsen theorem. 

The radial distribution $\rho_r(|z|)$ of an isotropic complex random variable $z=x+iy$ can be also recovered from the marginal distribution of its real part 
\begin{equation}
\rho_x(x)=\int\limits_{-\sqrt{R^2-x^2}}^{\sqrt{R^2-x^2}}\rho(|z|) dy=2\int\limits_{x}^{\infty}\frac{\rho(r) rdr}{\sqrt{r^2-x^2}}.
\end{equation}
Here $R$ is the radius of the support of the distribution, which can be infinite. The last integral is the Abel transform of the radial profile, which can be inverted via
\begin{equation}
\rho(r)=-\frac{1}{\pi}\int\limits_r^{\infty} \frac{d \rho_x(x)}{dx}\frac{dx}{\sqrt{x^2-r^2}},
\end{equation}
giving the exact one-to-one mapping between radial density profile of a complex number and the distribution of its real part.

One can ask whether a counterpart of this relation exists in the noncommutative free probability. Biely and Thurner~\cite{BielyThurner} conjectured that the Abel transform directly transfers to matrices. Later, it was pointed out that the Abel transform does not give a proper spectral density for the product of two GUE matrices~\cite{BurdaJanikWaclaw}. Recently, the authors has shown that such a procedure, coined as Abelization, works for normal matrices, the spectrum of which possesses an azimuthal symmetry~\cite{NowakTarnowskiLagged}. The biunitarily invariant matrices are in general not normal, which can be seen form the fact that the eigenvector correlation function $O(|z|)$ does not vanish.

Consider now the $R$-transform of the matrix $X+X^{\dagger}$, which is twice the Hermitian part of $X$.  This matrix appears in the description of the transient regime of a linear dynamical system~\cite{GRELA}.  Its cumulants can be related to the non-Hermitian cumulants, encoded in the determining sequence, via
\begin{equation}
R_{X+X^{\dagger}}(z)=\sum_{k=1}^{\infty}z^{k-1}\<\frac{1}{N}\tr (X+X^{\dagger})^{k}\>_c=\sum_{k=1}^{\infty}2z^{2k-1}\<\frac{1}{N}\tr(XX^{\dagger})^k\>_c=2zA_X(z^2).
\end{equation}
In the second equality we have used the fact that the only non-vanishing cumulants of the {\bf R}-diagonal operators are $\alpha_k=\<\frac{1}{N}\tr (XX^{\dagger})^{k}\>_c$, which can be extracted from the even powers of $X+X^{\dagger}$ exactly in two ways. The connected cumulants are encoded in the determining sequence $A$ instead of the $R$-transform of $XX^{\dagger}$, because in the expansion of the Green's function of $X+X^{\dagger}$ matrices $X$ and $X^{\dagger}$ are separated by either $\frac{\idm}{z}$ term or a plus sign.

\section{Applications of this formalism \label{sec:Applications}}
1. \textit{Quaternionic $R$ transform of Haar unitary matrix.}

We consider a unitary matrix $UU^{\dagger}=\idm$, the spectral density of which is uniform on the unit circle. Due to unitarity, $R_{UU^{\dagger}}(z)=1$ and from \eqref{eq:RelationRS} $S(z)=1$, hence $K(z)=1+z$. Substituting $z\to xA(x)$ and using \eqref{eq:AKrelation}, we obtain the quadratic equation
\begin{equation}
xA^2(x)+A(x)-1=0.
\end{equation}
Knowing that $A(0)=c_{1\bar{1}}=\<\frac{1}{N}\tr UU^{\dagger}\>_c=1$, we choose the appropriate branch of the solution. From \eqref{eq:QuatRTransform} we deduce the quaternionic R transform, which reads
\begin{equation}
\cR(Q)=\frac{1-\sqrt{1-4|w|^2}}{2|w|^2}\left(\begin{array}{cc}
0 & i\wb \\
i w & 0 
\end{array}\right).
\end{equation}
The same result was derived in~\cite{JaroszUnitary} using a different technique. We remark that the free cumulants are $\alpha_n=(-1)^{n-1} C_{n-1}$, where $C_{n-1}=\frac{1}{n+1} {{2n}\choose{n}}$ are the Catalan numbers, in agreement with~\cite{SpeicherUnitary}.

2. \textit{Free non-Hermitian Poisson.}

In Hermitian free probability, the counterpart of the Poisson distribution is the Wishart distribution, which has all cumulants the same, equal to $q$. The matrix model corresponding to this distribution is constructed as follows. Take $X$ a $N\times T$ rectangular matrix with iid  normal complex entries. The spectral density of $\frac{1}{T}XX^{\dagger}$ in the limit $N,T\to\infty$ with $T/N=q$ fixed is given by the Wishart law.

Let us consider now the isotropic ensemble with all $\alpha_n=\lim_{N\to\infty}\<\frac{1}{N}\tr (XX^{\dagger})^n\>_c=q$ the same for any $n\geq 1$ and vanishing all other mixed cumulants. The generating sequence is therefore $A(z)=\sum_{k=1}^{\infty}q z^{k-1}=\frac{q}{1-z}$. The $S$ transform reads $S(z)=[(1+z)(q+z)]^{-1}$ and the application of the Haagerup-Larsen theorem yields $F(r)=\frac{1-q+\sqrt{(q-1)^2+4r^2}}{2}$, where the positive branch of the solution was taken, so that the cdf is increasing. We easily recover the spectral density and the eigenvector correlator
\begin{eqnarray}
\rho(r)=\max(0,1-q)\delta^{(2)}(z)+\frac{1}{\pi \sqrt{(1-q)^2+4r^2}} \theta(\sqrt{q}-r), \\
O(r)=\frac{q \sqrt{(q-1)^2+4 r^2}-q^2+q-2 r^2}{2 \pi  r^2} \theta(\sqrt{q}-r).
\end{eqnarray}

Remarkably, the spectral density corresponds to the density of eigenvalues in the model $\frac{1}{T}XY^{\dagger}$, where $X$ and $Y$ are independent $N\times T$ matrices the entries of which are iid normal complex, considered in the limit $N,T\to\infty$, with $q=T/N$ fixed~\cite{GaussianProducts}, generalizing the result from the Hermitian case. The formula for the eigenvector correlation function is a new result.

3. \textit{Cumulants of products of Ginibre matrices.}

Let $X$ be a complex Ginibre matrix. The $S$ transform of $XX^{\dagger}$ reads $S(z)=(1+z)^{-1}$. Let us consider a product of $k$ independent Ginibre matrices. The multiplication law leads to $S_{k}=(1+z)^{-k}$. Using the relation \eqref{eq:SArelation} we obtain the algebraic equation for the determining sequence
\begin{equation}
(z A_k(z)+1)^{k-1}=A_k(z).
\end{equation}
The solution can be written in a power series~\cite{Knuth}
\begin{equation}
A_{k}(z)=\sum_{n=1}^{\infty}A_{n-1}(k-1,k-1) z^{n-1},
\end{equation}
where
 \begin{equation}
A_n(p,r)=\frac{r}{np+r}{{np+r}\choose{m}}=\frac{r}{n!}\prod\limits_{i=1}^{n-1}(mp+r-i)
\end{equation}
are the two parameter Fuss-Catalan numbers, also known as Raney numbers. The {\bf R}-diagonal cumulants are therefore $\alpha^{(k)}_n=A_{n-1}(k-1,k-1)=A_n(k,1)$. Such numbers has appeared in the free probability many times~\cite{Mlotkowski,Mlotkowski2,Mlotkowski3} and densities associated with them have been extensively studied~\cite{Forrester,ZyczkPenson}.

4. \textit{Commutator $[X,X^{\dagger}]$}

One of many measures of non-normality of a matrix (see e.g.~\cite{Elsner}) is defined through the spectral properties of a Hermitian matrix $C:=XX^{\dagger}-X^{\dagger}X$. Usually it is a square root of its Frobenius norm or the square root of the largest eigenvalue. Substituting the polar decomposition $X=PU$, one obtains $C=P^2+U(-P^2)U^{\dagger}$.
The unitary matrices assert that in the large $N$ limit the summands are free and their addition reduces to the addition of the corresponding $R$ transforms
\begin{equation}
R_C(z)=R_{P^2}(z)+R_{-P^2}(z)=R_{P^2}(z)-R_{P^2}(-z),
\end{equation}
since $R_{aX}(z)=aR_X(az)$.

In the simplest instance, the Ginibre matrix, the $R$-transform of the commutator reads $R_C(z)=\frac{z}{1-z^2}$, which corresponds to the distribution known as the Tetilla law, derived for the first time in~\cite{NicaSpeicherCommutator}. It was proven to be the limiting law for the anticommutator of Hermitian Wigner matrices~\cite{Deya}. Recently, it has also found an application in quantum information~\cite{ZyczkowskiTetilla}.

\section{Summary \label{sec:Summary}}
In this work, we have formulated a diagrammatic construction  for {\bf R}-diagonal operators or, equivalently,  bi-unitary invariant random matrices in the limit when the size of the matrix  tends to infinity. Relations  between individual cumulants and moments were known, but were expressed usually in terms of  zeta functions and their inverses (M\"{o}bius functions), and the explicit functional relation between the corresponding generating functions was unknown. 
Providing such an operational construction is the first new result of this work.  Then, adapting  the formalism of free random variables for  the Hermitian and the non-Hermitian ensembles to the case of {\bf R}-diagonal operators,  we have obtained a concise proof of the original Haagerup-Larsen theorem  for the isotropic spectra (single ring theorem) and of its recent extension for the case of the correlation function, which involves overlaps of left and right eigenvectors of nonnormal matrices.   

Hitherto, all proofs of the single ring theorem in the formalism of free random variables were based on the analytic methods, so providing the concise diagrammatic proof  is  the next new result. It is also interesting to speculate, why the eigenvector part of single ring theorem was missed for almost 20 years (counting from original formulation by Feinberg  and Zee), despite several follow-ups and generalizations  in physical  and mathematical literature.   We dare to link this fact to the very subtle and underappreciated role of the complex parameter $w$ in the "electrostatic potential"  (or the regularized Fuglede-Kadison  determinant  in  mathematical language) in (\ref{FugKad}).  When considering the spectra, this parameter serves solely as the infinitesimal regulator, which  is  put to zero after performing the average over the ensemble.  However, since it appears  also in a non-trivial way in off-diagonal components of the quaternionic Greens' function (\ref{eq:GComponents}), contributing to the left-right eigenvector correlations, particular care has to be taken during the limiting procedure.  Actually, looking at our prescription for R-transform for bi-unitarily invariant ensembles~(\ref{eq:QuatRTransform}) one clearly sees 
that taking the limit $|w| \rightarrow 0$ in a haste way makes our  construction meaningless. Moreover, keeping $w\neq 0$ causes that $\cQ$ do not commute with the linearized matrix $\cX$, which in turn is crucial in encoding all mixed cumulant in the quaternionic $R$-transform.

Last but not least, taking into account the rapidly  growing impact  of free random variable calculus onto so many branches of modern applications~\cite{MIMO,QUANTUM,STABILITY} 
we hope that the presented here operational construction will contribute  to further interweaving of both communities of mathematicians and practitioners.

\section*{Acknowledgments}
WT is grateful to Professor Marek Bo\.{z}ejko for fruitful discussions and for posing the problem of the non-Hermitian free Poisson, which initiated this work. The authors appreciate also the discussions with  Roland Speicher. The research was supported by the MAESTRO DEC-2011/02/A/ST1/00119 grant of the National Center of Science. WT also appreciates the financial support from the Polish Ministry of Science and Higher Education through ‘Diamond Grant’ 0225/DIA/2015/44 and the scholarship of Marian Smoluchowski Research
Consortium Matter Energy Future from KNOW funding.

\end{document}